# Context Protecting Privacy Preservation in Ubiquitous Computing

Arijit Ukil
Innovation Labs
Tata Consultancy Services
Kolkata, India
arijit.ukil@tcs.com

*Abstract—* In ubiquitous computing domain context awareness is an important issue. So, in ubiquitous computing, mere protection of message confidentiality is not sufficient for most of the applications where context-awareness can lead to near deterministic ideas. An adversary might deduce sensitive information by observing the contextual data, which when correlated with prior information about the people and the physical locations that are being monitored by a set of sensors can reveal most of the sensitive information. So, it is obvious that for security and privacy preservation in ubiquitous computing context protection is of equal importance. In this paper, we propose a scheme which provides two layer privacy protection of user's or application's context data. Our proposed context protecting privacy preservation scheme focuses on protecting spatial and temporal contextual information. We consider the communication part of ubiquitous computing consists of tiny sensor nodes forming Wireless Sensor Networks (WSNs). Through simulation we show the efficacy of our scheme. We also demonstrate the capability of our scheme to overcome the constraints of WSNs.

*Keywords-ubiquitous computing; context-awareness; privacy preservation; key management;*

## I. INTRODUCTION

Ubiquitous computing enhances computer use by making many computers available throughout the physical environment, while making them effectively invisible to the user. This requires functioning of multitude of devices in the environment to be oblivious to the users. Mark Weiser in his paper "The Computer for the Twenty-First Century" defines ubiquitous computing as a technology that "weave itself into the fabric of everyday life until it is indistinguishable from the it" [1]. This point of view leads to the notion that almost everything in the fabric of ubiquity points to context-awareness. Context-awareness evolved from the idea to express the ability of systems and components to respond to the situation of the user that interacted with this system. Research in context-aware computing has made numerous attempts to model not only human attributes and behavior but also how we relate to our environment [2]. Ubiquitous applications requires continuous monitoring, gathers vast amounts of sensitive electronic information about the users, which is the basis of finding opportunities for data interception, theft and surveillance. For example, the disclosure of both spatial and temporal data through traffic analysis, may allow tracking the relative or actual information through correlation with prior knowledge. With so much retrievable personal information available through Internet, it is now becoming difficult and challenging to protect privacy. Good amount of research effort has been gone into the research of privacy preservation, particularly in ubiquitous domain, where distributed computing poses a great threat. In [3], it is observed that ubiquitous computing environments require security and privacy architecture based on trust rather than just user authentication and access control. Burnside et al. [4] described a resource discovery and communication system designed for security and privacy.

So, we can observe that along with user privacy protection by hiding users' identity or location, we require to protect data privacy. In a privacy-preserving data aggregation (PPDA) protocol, sensor data are partially exposed to neighboring trusted sensor nodes so that data aggregation can be achieved on the way to the source node without revealing the actual data to the trusted sensor nodes or adversaries [5-7]. In his famous paper [8], Yao has introduces the millionaire problem, which can be summarized as: A and B are two millionaires who want to find out who is richer without revealing the precise amount of their wealth. The objective of privacy preserving data mining is to meet the required privacy requirements and to provide data mining outcome [9]. There are number of research proposals and algorithms exist for solving the stated problem. In [10], the problem of privacy preservation in a peer-to-peer network application is addressed. In [11], Wenbo He et.al. propose schemes to achieve data aggregation while preserving privacy. The scheme they proposed, CPDA (Cluster-based Private Data Aggregation) performs privacy-preserving data aggregation in low communication overhead with high computational overhead. In CPDA, each cluster leverages the additive property of polynomials to calculate the desired aggregate value.

From this background, we propose our context protecting privacy preservation scheme. This scheme has two layers. In first layer, the contextual information derived from spatial and temporal domain identity of the user is protected. In second layer, the actual contextual data privacy protection is made using the concept of PPDA.

The paper is organized as follows. In Section 2, we illustrate the system architecture. In Section 3, we present first layer of our scheme, where location and timing data privacy is protected. In Section 4, we describe the privacy preservation method of the contextual data itself by using PPDA. In Section 5, we present the simulation result and analysis. Lastly, we conclude the paper in Section 6 with conclusion and future scope of work.

## II. SYSTEM MODEL

In this section, we illustrate the system model, based on which our context protecting privacy preservation scheme is based. We consider N number of sensor nodes are present in the distributed network, which is a part of overall ubiquitous computing system. We take Home Gateway (HG) as the central server which is connected to the Internet. The sensor nodes are bi-directional and they have single-hop (direct) or multi-hop link with the HG in order to maintain connectivity with the outside world. For illustration purpose we have taken N= 8. This is shown in Fig.1.

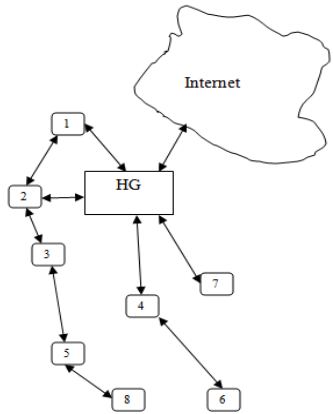

Figure 1. Home Gateway based ubiquitous computing architecture

The objective of our scheme is that the contextual information regarding any of the sensor nodes ($s_i$, $i \in N$) need to be protected. These information are $s_{is}$, $s_{it}$, which are for spatial and temporal identity of the nodes. Apart from that it is required that even in the case these data are revealed to an attacker, he/she cannot make out the content of the data. This is in fact, a doubly locked privacy preservation scheme, where both the locks (schemes) are independent. From functional point of view, contextual data has two parts. First part, it is being made anonymous so that even in the case actual data is revealed, source or destination turns out to be vague. The second part consists of preserving the privacy of the actual data itself by data perturbation technique so that attacker gets confused. This is shown in Fig. 2. Our proposed scheme enables users to control their personal data. If the user does not require concealing its contextual information, then the system bypasses the scheme and directly delivers the information in traditional way. But in the case, the user wants its contextual data to be privacy protected; the user needs to inform the system about the amount of protection it requires. If only anonymity is sufficient, then second layer is ignored. If data privacy preservation or data perturbation based PPDA is required then first layer is bypassed. If the user wants absolute privacy protection, then both the layers are used for its contextual data protection. This algorithm is shown in Fig. 3.

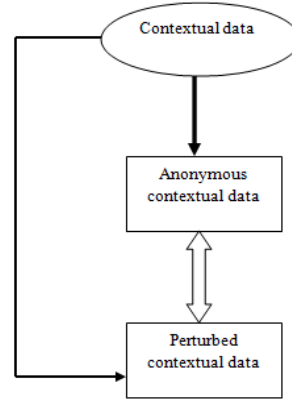

Figure 2. Functional model

## III. ANONYMITY OF CONTEXTUAL INFORMATION

In this section, we propose our scheme of contextual data privacy protection by anonymity. In this case, we consider only spatial and temporal privacy. We would like to remark that the classification of context-oriented privacy into the two categories of spatial and temporal privacy only reflects the current state-of-the art, and should not be treated as a comprehensive classification. In order to illustrate this, we present our scheme for location privacy, which can be extended to temporal privacy by slight modification.

In order to conceal the location or spatial information of a node, we need to protect the poor privacy protection performance of conventional routing protocols. To achieve our objective of location privacy protection, we consider the phantom routing strategy [12]. In this scheme, the source location privacy is protected through directing the periodic messages from the source node towards different paths in the network. This prohibits the malicious nodes or the attacker from receiving a stable stream of messages that would enable back-tracing the source. Instead of that, by the received messages the attacker is led towards phantom sources. This routing strategy is depicted in Fig. 4. In this scheme, there are two phases exist for packet delivery. First one, which originates from the source ($S$) is a pure or directed random walk for a given number of hops that directs the message to a phantom source or flooding source F away from the original source S. The second phase, which is message flooding phase delivers the message to the destination D. In this case random walk at the first phase leads to different flooding node, which makes tracing back more difficult. If the malicious node M detects a message forwarded by node F and moves to that node to get closer to the source, the next message is unlikely to follow the same random path. This makes M's previous move worthless. For more protection, we can apply greedy random walk approach, which is a two way random walk, performed both from the source and the source. It is inspired by the observation that if M gains a good coverage of the network by distributing a number of observation points around the source, the source location could be approximated because the flooding phase would reveal too much information. In order to avoid this, instead of using flooding to deliver the message to the destination D, the destination node sets up a random walk which serves as the receptor of the messages. Each message is randomly forwarded from a source until it reaches the receptor, and is then forwarded to the destination through the pre-established path. A further advantage that the random walk offers is that the safety period improves as the network size increases, as the paths followed by subsequent messages, and consequently the malicious nodes, become more diverse. The diversity of the paths is not, however, the only issue that the random walk implementation needs to ensure. The main purpose of this phase is to send each message to a phantom source that is far from the original source. The second problem is finding the flooding zone such that it is as minimum (with respect to number nodes message needs to be flooded to) as possible which will make trace back low probabilistic. Finding this optimality condition can be explained by an example.

Let, probability of original source detection be $P_r$, which is a very small number, $P_r \rightarrow 0$. Typically, $P_r < 10^{-2}$. In order to find the destination (assuming that our first phase has been broken. Source S and flooding node F are identified) D, the attacker has to try out each of the flooded message from F. Now, if N number of nodes are available in the flooding zone and average H number of hops possible in that zone, then message is broadcast to K number of nodes, where

$$K = \frac{N!}{H!(N-H)!}$$

So, finding S from K possible nodes is of probability: 1/K. So, the optimal condition is:

$$P_r < \frac{1}{K} \rightarrow K > \frac{1}{P_r}$$

Now, let us consider it numerically for practical case. Consider,

$$P_r = 10^{-2}$$
$$H = 3$$

N turns out to be = 10 to satisfy the condition: $K > \frac{1}{P_r}$.

This is very nominal. In the case the average hop number is 4, N is approximately 8. So, we observe that, very low probability of trace-back condition can be achieved with few numbers of nodes selected in the flooding zone. The proposed scheme when compared to basic flooding, the energy consumption, which mainly depends on the number of the transmitted messages, is not increased. The message latency, however, could be significantly increased depending on the length of the random walk. This may not be a problem as most of the current day's privacy preservation applications are non real-time in nature.

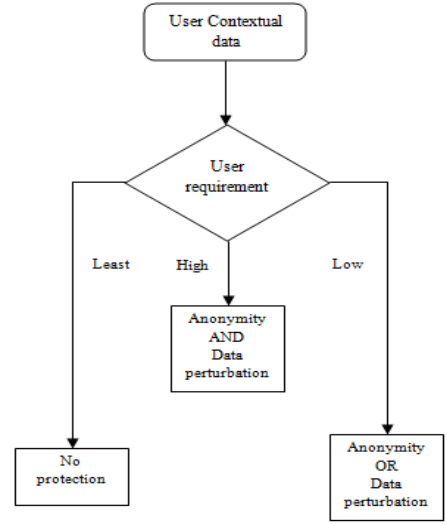

Figure 3. Functional model

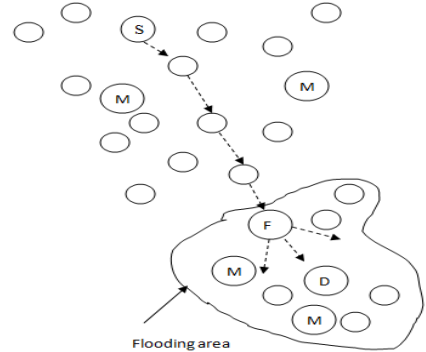

Figure 4. Context anonymity routing

IV. PERTURBATION OF CONTEXTUAL INFORMATION

In this section, we present the scheme for privacy preservation of user data. The objective is similar that of [8], i.e., we consider a scenario where data aggregation needs to be done in privacy-preserved way for distributed computing platform. There are number of data sources which collect or produce data. The data collected or produced by the sources is private and the owner or the source does not like to reveal the content of the data. But the collected data from the source is to be aggregated by an aggregator, which may be a third party or part of the network, where the data sources belong. The data sources do not trust the aggregator. So the data needs to be secure and privacy protected. In tune of that, we propose a scheme which is secure and privacy preserved. The computation for the aggregation is based on the concept of Secure Multiparty Computation (SMC) [13]. In this case, we need to slightly modify the routing process initialization from the source. Instead of one, we consider two sources (S1 and S2) and one Aggregator-Forwarder (AF) node, where the data of the source nodes will aggregate. The AF node aggregates the data of S1 and S2 and forwards the aggregated value towards the Flooding node. Aggregation process is governed by data perturbation technique, where the AF cannot find out the exact content of the data of S1 and S2. This is depicted in Fig. 5. Here, we follow the scheme proposed in [7] by He et al. The scheme they proposed, CPDA (Cluster-based Private Data Aggregation) performs privacy-preserving data aggregation in low communication overhead with high computational overhead. They have assumed a self-organized multi-hop wireless sensor networks. The scheme CPDA though very much effective, but suffers from two critical limitations:

1. Computation of the privacy preservation algorithm increases exponentially with the number of source nodes. In fact, its computational complexity is $O(N^2)$, where N= number of nodes.
2. In most of the practical scenarios, the source nodes cannot communicate directly with each other or peer-to-peer. In these cases, CPDA is useless.

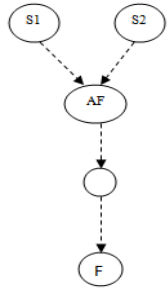

Figure 5. Privacy routing

Our proposed scheme consists of three parts:
1. Key management
2. Data value distortion
3. Data aggregation

*A. Key Management:*

In order to accomplish these objectives, we first form cluster of the source nodes. Let, there be N number of source nodes and each cluster consists of n number of source nodes. So, there will be N/n number of clusters. The key management process starts by key pre-distribution stage. In the pre-distribution phase, a large key-pool of K keys and corresponding identities are generated. These K number of keys are divided into two banks. One bank consists of k number of keys, which is used for source node's communication with AF. The rest K-k number of keys form another bank which are required for communication between S1 and S2 via aggregator node. So, the key management scheme consists of two parts:

*Source to AF:* Each source node has K-k number of keys shared with AF. As, all the source nodes possess the same keys, it is totally unsecure when a source node communicates with AF node with the shared key. Any malicious source node can decipher the source nodes' communication with AF and can launch attack very easily. In order to avoid this, in the pre-distribution phase, the source-AF key bank is randomly permuted and reordered for each source-AF pair. This ordering of the key bank is stored in the AF for each source. Now, the source node communicates with AF through one of its shared keys. To accomplish this action, the source node first generates a random number between 1 and K-k. This random number (Rc) is sent to AF in plain text. AF understands that the source node will encrypt the next message by the Rcth number key of the key bank.

*Source to Source (S1 to S2):* It is assumed that source to source direct communication does not exist and this has to happen securely through AF. In order to achieve that, the k number of keys are stored in the source nodes, which AF is unaware of. It is also a requirement that other source nodes should not decipher the message source 1 sends to source 2. As the k keys are same for all the source nodes, it becomes easy for another source node to decrypt the plain text, i.e. source 3 can decrypt what source node 1 and source node 2 are communicating. To avoid this situation, source node 1 and source node 2 separately permute the key bank order of the k number of keys dedicated for source-source communication and reorder that randomly. After that, they pass the permute function to each through AF using their pair-wise key with the AF.

*B. Data value distortion:*

Let the data values at S1 and S2 be x and y, z be the dummy variable at the aggregator (A). In the first step, the aggregator sends three seeds a,b and c to S1 and S2. Based on that A computes

$$\alpha_{S1}^A = z + R_1^A a + R_2^A a^2$$
$$\alpha_{S2}^A = z + R_1^A b + R_2^A b^2$$
$$\alpha_A^A = z + R_1^A c + R_2^A c^2$$

Where $R_1^A$ and $R_2^B$ are two random numbers generated by A. Similarly, S1 computes

$$\alpha_{S1}^{S1} = x + R_1^{S1} b + R_2^{S1} b^2$$
$$\alpha_A^{S1} = x + R_1^{S1} a + R_2^{S1} a^2$$
$$\alpha_{S2}^{S1} = x + R_1^{S1} c + R_2^{S1} c^2$$

Similarly S2 computes

$$\alpha_A^{S2} = y + R_1^{S2} a + R_2^{S2} a^2$$
$$\alpha_{S1}^{S2} = y + R_1^{S2} b + R_2^{S2} b^2$$
$$\alpha_{S2}^{S2} = y + R_1^{S2} c + R_2^{S2} c^2$$

Where $R_1^{S1}$ and $R_2^{S1}$ are two random numbers generated by S1, $R_1^{S2}$ and $R_2^{S2}$ are two random numbers generated by S2. After that, the calculated, $\alpha_{S1}^A$ and $\alpha_{S2}^A$ are sent to S1 and S2 by A, securely as described earlier. Similarly, $\alpha_A^{S1}$ and $\alpha_{S2}^{S1}$ are sent to source node 2 and A by S1 and $\alpha_A^{S2}$ and $\alpha_A^{S2}$ and $\alpha_{S1}^{S2}$ are sent to A and S1 by S2.

*C. Data aggregation:*

After the private data values (x and y) are distorted, all the nodes aggregates the values available to them and generates aggregated result. S1 calculates , S2 calculates and A calculates .

Where,
. These aggregated results from S1 and S2 are securely sent to the aggregator A. Now, the aggregator has the simple task to solve the above equation for (x+y+z) with the knowledge of the values of a,b,c and , and . After solving for D = x+y+z, node A knows its data z, so it can find out the result (x+y).

## V. SIMULATION RESULTS

In this section, we show a comparative study between our proposed scheme and the CPDA scheme in [11]. The objective of our work is to find a simpler, efficient privacy preserved data aggregation scheme, which has scalability and can be highly effective in some practical scenario like discussed in the motivation section. From that perspective, we see that computation time requirement to run SPPDA comes out to be around 1 msec in an Intel Core 2 duo PC with CPU speed 3 GHz and RAM of 2 GB. Whereas, if we increase the number of source nodes to 3, the overall computation speed becomes 3 msec as shown in fig. 6. As in our scheme, in most cases, there will be fixed two number of source nodes, the computational time becomes fixed. This is indeed a necessary requirement if the overall system is real-time in nature and for resource limited sensor nodes in WSN.

In CPDA scheme, there exists certain probability where private data may be disclosed. This can only happen when the source nodes exchange messages within the cluster. This can be estimated as:

Where Dmax = maximum cluster size, pc = minimum cluster size (= 3, two source nodes and one aggregator), k = cluster size, b = probability that link level privacy is broken, P(k=m) = probability that a cluster size is m. In our case, pc = Dmax = k = 3, P(k=m) =1. So, we have plotted P(b) for CPDA and our scheme in fig. 7. It is observed that the probability of privacy compromised in CPDA has much steeper slope.

In CPDA, a requirement is that a pair of source nodes possessing same pair of keys, where the keys are taken randomly from a large pool of key, should be high. Otherwise, the scheme cannot work. But, this requirement helps malicious nodes to capture at least some of the communication, if it has common pair of keys. This probability also increases with number of source nodes increase when total number of keys in the key pool is constant. In our scheme, there is no requirement like that.

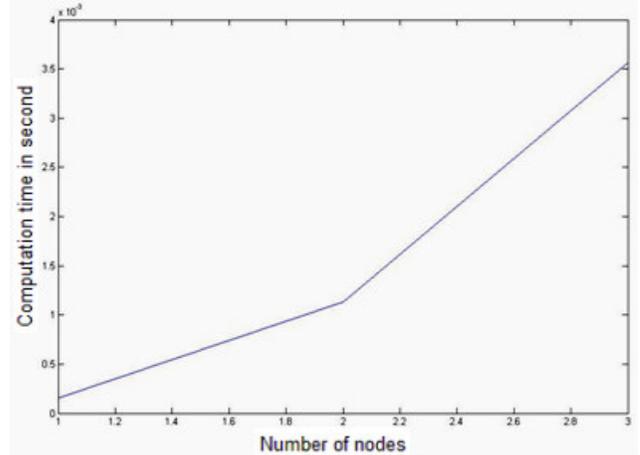

Figure. 6. Computation time requirement

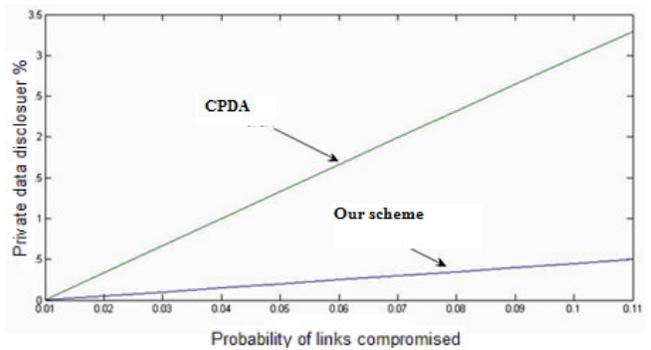

Figure. 7. Probability of private data disclosure

## VI. CONCLUSION

In this paper, we proposed a scheme which aims to protect the privacy of the contextual data mainly in ubiquitous computing environment. It is a two-tier scheme and user has the choice of opting for none, both or any of the tiers as per its security and privacy requirements. With growing number of ubiquitous applications like Home Gateways developed, it becomes very important to conceal one's contextual information, which indirectly can destabilize the security policies. Our scheme, in that sense is a very important development. Our future work will mainly focus on the complexity analysis and actual implementation in real test-bed.

REFERENCES

[1] Mark Weiser, "The Computer for the Twenty First Century," Scientific American, pp. 94-104, September, 1991.
[2] R. da Rocha and M. Endler, "Evolutionary and efficient context management in heterogeneous environments," In Proceedings of the 3rd International Workshop on Middleware for Pervasive and Ad-hoc Computing, pp. 1-7, 2005.

[3] L. Kagal, T. Finin, and A. Joshi, "Trust-Based Security in Pervasive Computing Environments," IEEE Computer, vol. 34, no. 12, pp. 154 – 157, 2001.

[4] M. Burnside, D. Clarke, Mills, A. Maywah, S. Devadas, R. Rivest, "Proxy-Based Security Protocols in Networked Mobile Devices", 17th ACM Symp. on Applied Computing, pp. 265–272, 2002.

[5] J. Yao, G. Wen, "Protecting classification privacy data aggregation in wireless sensor networks," In Proceedings of the 4th International Conference on Wireless Communication, Networking and Mobile Computing, WiCOM, Dalian, pp. 1–5, 2008.

[6] W.S. Zhang, C. Wang, and T.M. Feng, "GP2S: Generic privacy-preservation solutions for approximate aggregation of sensor data, concise contribution," In Proceedings of the 6th Annual IEEE International Conference on Pervasive Computing and Communications, pp.179–184, 2008.

[7] G. Taban and V.D. Gligor, "Privacy-preserving integrity-assured data aggregation in sensor networks," In Proceeding of International Symposium on Secure Computing, pp. 168–175, 2009.

[8] A. Yao, "Protocols for secure computations," Proceedings of the 23rd Annual Symposium on Foundations of Computer Science, pp.160-164, 1982.

[9] S.R.M. Oliveira and O.R. Zaïane, "Achieving Privacy Preservation when Sharing Data for Clustering," Springer LNCS, 3178, pp. 67-82, 2004.

[10] Q. Huang, H.J. Wang, and N. Borisov, "Privacy-preserving friends troubleshooting network," Symposium on Network and Distributed Systems Security (NDSS), pp. 184-194, 2005.

[11] W. He, et al., "PDA: Privacy-preserving Data Aggregation in Wireless Sensor Networks," IEEE INFOCOM, pp. 2045 – 2053, 2007.

[12] C. Ozturk, Y. Zhang, and W. Trappe, "Source-location privacy in energy-constrained sensor network routing," 2nd ACM workshop on Security of ad hoc and sensor networks, pp. 88–93, 2004.

[13] S. Goldwasser, "Multi-party computations: Past and present," 16th Annual ACM symposium on Principles of distributed computing, 1997.